\documentclass[aip,reprint]{revtex4-1}

\usepackage{graphicx}
\usepackage{dcolumn}
\usepackage{bm} 

\draft 

\usepackage{fancyhdr}
\begin{document}


\title{Effective uniaxial anisotropy in easy-plane materials through nanostructuring} 



\author{J. Fischbacher}
\author{A. Kovacs}
\author{H. Oezelt}
\author{M. Gusenbauer}
\affiliation{Center for Integrated Sensor Systems, Danube University Krems, Viktor Kaplan Str. 2/E, 2700 Wiener Neustadt, Austria}
\author{D. Suess}
\affiliation{Christian Doppler Laboratory for Advanced Magnetic Sensing and Materials, University of Vienna, W{\"a}hringer Str. 17, 1090~Wien, Austria}
\affiliation{Faculty of Physics, University of Vienna, Boltzmanngasse 5, 1090~Wien, Austria}
\author{T. Schrefl}
\email[Electronic mail: ]{tschrefl@gmail.com}
\affiliation{Center for Integrated Sensor Systems, Danube University Krems, Viktor Kaplan Str. 2/E, 2700 Wiener Neustadt, Austria}


\date{\today}

\begin{abstract}
Permanent magnet materials require a high uniaxial magneto-crystalline anisotropy. Exchange coupling between
small crystallites with easy-plane anisotropy induces an effective uniaxial anisotropy if arranged accordingly. Nanostructuring
of materials with easy-plane anisotropy is an alternative way to create hard-magnetic materials. 
The coercivity increases with decreasing feature size. The resulting coercive field is about 12 percent of the anisotropy field for a crystal 
size of 3.4 times the Bloch parameter.
\end{abstract}


\maketitle 


\thispagestyle{fancy}

The growing demand for permanent magnets in energy conversion \cite{gutfleisch_magnetic_2011} triggered the search for new permanent-magnet materials \cite{drebov_ab_2013}. Wind power as well as hybrid and electric vehicles require high-performance permanent magnets. A prerequisite for a permanent-magnetic phase is a high uniaxial magneto-crystalline anisotropy. In a material with uniaxial anisotropy work is required to rotate the magnetization out of the easy axis. When made of materials with a sufficiently high \cite{skomski_magnetic_2016} magneto-crystalline anisotropy, the magnet can resist demagnetization by the self-demagnetizing field plus an additional external field up to the nucleation field. Several magnetic phases with a high magneto-crystalline anisotropy are easy-plane materials. The magnetization lies preferable in a plane perpendicular to a crystal symmetry axis. One way to make an easy-plane material uniaxial is through interstitial atoms which expand the lattice. A prominent example for this approach is Sm$_2$Fe$_{17}$N$_3$ where the anisotropy constant changes from $K = -0.8$~MJ/m$^3$ (easy-plane Sm$_2$Fe$_{17}$) to $K=7.1$~MJ/m$^3$ (uniaxial Sm$_2$Fe$_{17}$N$_3$). \cite{brennan_anisotropy_1995} Possible candidate materials for a rare-earth-free permanent magnet are FeSn based alloys. \cite{goll_magnetic_2015} Whereas Fe$_3$Sn shows an easy-plane anisotropy, ab initio simulations show that substituting Sn by Sb may lead to uniaxial anisotropy. \cite{sales_ferromagnetism_2015}

\begin{figure*}
\includegraphics{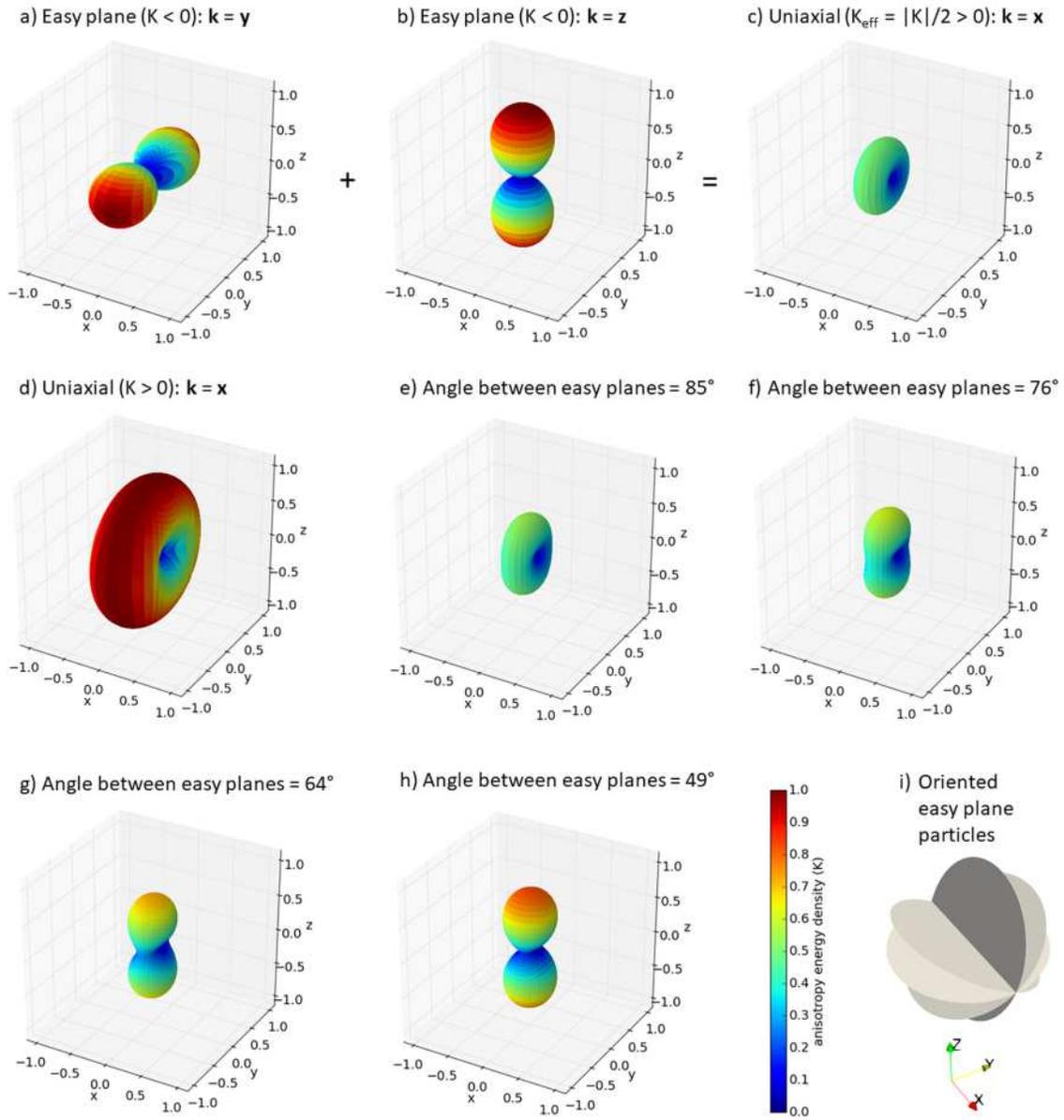}
\caption{\label{fig_ani} Anisotropy energy density as function of direction. The average of an easy-plane anisotropy in $xz$ (a) and an easy plane in $xy$ (b) gives an effective uniaxial anistropy in $x$ direction (c). (d): uniaxial anisotropy in $x$ direction. (e) to (h): average of two easy-plane anisotropies with different angles between the planes. (i): Sketch of oriented easy-plane particles \cite{balasubramanian_high-coercivity_2016}.}
\end{figure*}

Skomski et al. \cite{skomski2009graded} showed another way to create a uniaxial magnet from easy-plane materials. In Figure 4 of their work on nanostructured permanent magnets \cite{skomski2009graded} they illustrate the concept of hard-magnetic soft-soft composites. When exchange coupled at 90 degrees between their easy planes, two grains show uniaxial anisotropy with an effective uniaxial anisotropy constant $K_{\mathrm{eff}} = |K|/2$. An experimental realization of a hard-magnetic soft-soft composite was reported by Balasubramanian et al. \cite{balasubramanian_high-coercivity_2016}. However, the nanostructure is more complex. Instead of a 90 degrees relation between the $c$ axes of neighboring grains, the particles are oriented so that all easy planes are parallel to a common axis (see Fig.~\ref{fig_ani}i). In such clusters of aligned easy-plane Co$_3$Si particles with a size of 10 nm they measured coercivities of $\mu_0 H_\mathrm{c} = 1.74$~T  and $\mu_0 H_\mathrm{c} = 0.37$~T at 10 K and 300 K, respectively. The high coercivity of the magnet was attributed to exchange interactions which form an effective anisotropy. 

Nanostructures made of easy-plane materials may have the potential to be used as new permanent-magnet materials. There are two possible niches for new hard-magnetic phases on the market: (1) Cheap magnets with an intermediate energy density product in between ferrites ($< 38$~kJ/m$^{3}$) and Nd-Fe-B ($> 200$~kJ/m$^{3}$) \cite{coey2012permanent}. (2) Magnets with an energy density product comparable with Nd-Fe-B but without or reduced heavy-rare-earth content \cite{hirosawa2017perspectives}. Although, we used the intrinsic magnetic properties for Fe$_3$Sn for most of the simulations presented in this work, the results have general validity. In Fig.~\ref{fig_hc}, which summarizes the main results, the expected intrinsic coercive field (in units of the anisotropy field) is given as function of the feature size (in units of the Bloch parameter). Provided proper nanostructuring is possible, Fe$_3$Sn may fill the gap between ferrites and Nd-Fe-B with an estimated energy barrier of 100 kJ/m$^{3}$. On the other hand, nanostructuring Sm$_{2}$Fe$_{14}$B, which has an anisotropy constant of $K = -12$~MJ/m$^{3}$,\cite{sagawa1985magnetic} may give a magnet with an estimated coercive field $\mu_{0}H_{\mathrm c} > 1.5$~T and an energy density product $BH_\mathrm{max} > 200$~kJ/m$^{3}$. The magnetization of both materials is sufficiently high so that bonded magnets made of structured particles will still show good magnetic properties. 

We use micromagnetics to compute the effective uniaxial anisotropy resulting from exchange interactions between crystals with easy-plane magneto-crystalline anisotropy. We investigate the influence of the relative orientation of the particles and the particle size. When magnetic particles are small compared to the exchange length, their local anisotropies can be averaged to find an effective anisotropy. Herzer et al.\cite{herzer2005} used this approach in order to describe the properties of nanocrystalline soft magnets with the random-anisotropy model. In exchange-spring hard magnets the effective anisotropy resulting from the exchange interactions between a magnetically hard and a magnetically soft phase is given by the average anisotropy constant of both phases \cite{skomski1993giant} if the lateral extension of the phases is small enough. Here we apply a similar idea in order to investigate the influence of the relative orientation of two easy-plane crystallites on the average anisotropy. To investigate the influence of the crystalline size on the coercivity we apply finite element micromagnetic simulations.

\begin{figure}
\includegraphics{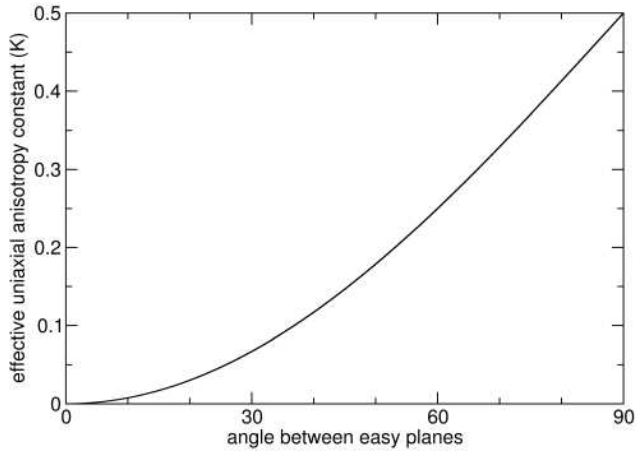}
\caption{\label{fig_keff} Two strongly exchange coupled particles with easy-plane anisotropy show an effective uniaxial anisotropy. The plot shows the effective anisotropy constant as function of the angle between the easy planes.}
\end{figure}  

\begin{figure}
\includegraphics{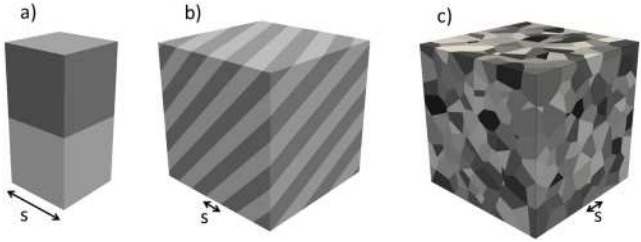}
\caption{\label{fig_models} Nanostructures used for micromagnetic simulations. (a) two crystallites, (b) grain with multiple twins, (c) cluster of 1000 particles. In (a) and (b) the angle between neighboring crystals was either 90, 85 or 76 degrees, in (c) the c-axes are randomly oriented in plane. The feature size, $s$, is varied from $0.85  \delta_0$ to $8.5 \delta_0$.}
\end{figure}  

\begin{figure}
\includegraphics{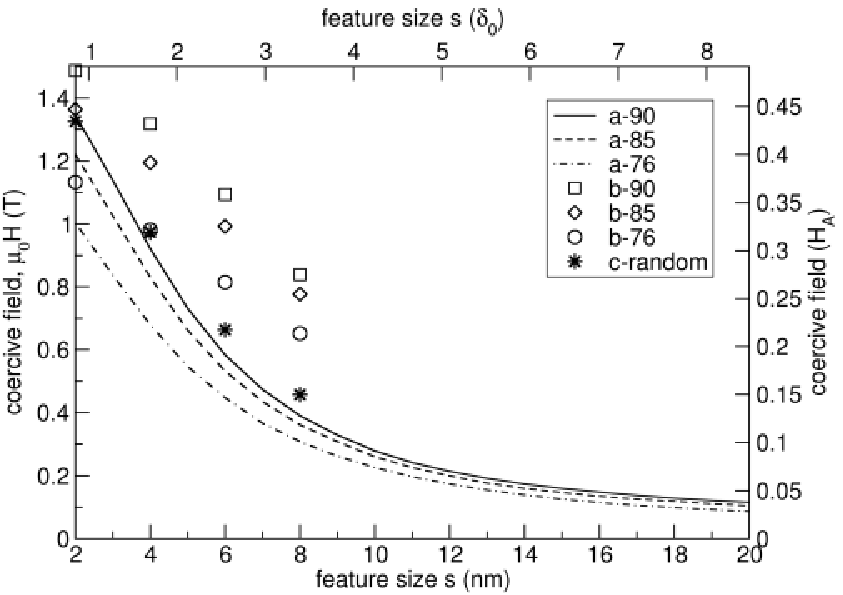}
\caption{\label{fig_hc} Computed coercive field as function of particle size for two exchange coupled crystals at different angles. The letters give the structure as shown in Fig.~\ref{fig_models}. The numbers denote the angle between neighboring crystals.}
\end{figure}

Let us consider a magnetic particle with a magneto-crystalline anisotropy constant $K$. Its magneto-crystalline anisotropy energy density is given by $K\sin^{2}{\varphi}$ where $\varphi$ is the angle between the magnetization and an axis $\bm{k}$. If $K > 0$ we have to perform work to rotate the magnetization out of $\bm{k }$, which is the easy axis. Fig.~\ref{fig_ani}d shows the work required to rotate the magnetization in a certain direction for a particle with the easy axis parallel to $\bm{x}$. If $K < 0$ we have to perform work  to rotate the magnetization out of a plane perpendicular to $\bm{k}$. Fig.~\ref{fig_ani}a shows the work required to rotate the magnetization out of the $xz$-plane, which is the easy plane. Here $\bm{k}$ is parallel to $\bm{y}$. In  Fig.~\ref{fig_ani}b  $K < 0$ and $\bm{k}$ is parallel to $\bm{z}$. The easy plane is rotated by 90 degrees with respect to the configuration of Fig.~\ref{fig_ani}a. Averaging the anisotropy energy densities of the particles in Fig.~\ref{fig_ani}a and Fig.~\ref{fig_ani}b gives the anisotropy energy density shown in Fig.~\ref{fig_ani}c. The effective anisotropy of two exchange coupled particles with easy-plane anisotropy is uniaxial with $\bm{k}$ parallel to the intersection line of the planes if the angle between the planes is 90 degrees. From the comparison of Fig.~\ref{fig_ani}c (average of easy-plane anisotropies at 90 degrees) and Fig.~\ref{fig_ani}d (uniaxial anisotropy) we see that the effective anisotropy constant $K_\mathrm{eff} = K/2$. However, the effective uniaxial anisotropy decreases with increasing deviation from the perpendicular arrangement between the particles as shown in Fig.~\ref{fig_ani}e to Fig.~\ref{fig_ani}h. 

The effective uniaxial anisotropy for two strongly coupled particles with
angle $\theta$ between their easy planes has been derived by Skomski \cite{skomski2016permanent}. He showed that the effective easy axis is $\mathbf{k} = \mathbf{n}_{1} \times \mathbf{n}_{2}$, where $\mathbf{n}_{1}$ and $\mathbf{n}_{2}$ are the normal vectors to the easy planes. He obtained $K_{\mathrm{eff}} = |K|(1-\cos \theta)/2$ for the effective anisotropy constant. Here we derive the effective anisotropy using basic Calculus. In Fig.~\ref{fig_ani}e to Fig.~\ref{fig_ani}h the symmetry axis, $\bf{k}$, of each particle lies in the $yz$-plane. The unit vectors parallel to the symmetry axes of the particles are ${\bf k}_{1}=(0,-\sin(\theta/2),\cos(\theta/2))$ and ${\bf k}_{2}=(0,\sin(\theta/2),\cos(\theta/2))$. When $\bf{m}$ is the unit vector of the magnetization, the average anisotropy energy density of the two particles is $e = (-K({\bf m}\cdot{\bf k}_{1})^{2}-K({\bf m}\cdot{\bf k}_{2})^{2})/2.$ Here we used $\sin^{2}\varphi = 1 - \cos^{2}\varphi$ and dropped constant terms. 

We write the unit vector of the magnetization in polar coordinates \\ $\bf{m} = (\cos\varphi_{\mathrm m} \sin\theta_{\mathrm m},\sin\varphi_{\mathrm m} \sin\theta_{\mathrm m},\cos\theta_{\mathrm m})$. With the vectors $\bf{m}$, ${\bf k}_{1}$, and ${\bf k}_{2}$ as defined above we can write the anisotropy energy density
\begin{eqnarray}
\label{eq_av}
e(\varphi_{\mathrm m},\theta_{\mathrm m}) & =  - K & 
     \left( \sin^{2}{\left (\frac{\theta}{2} \right )} \sin^{2}{\left (\varphi_{\mathrm m} \right )} \sin^{2}{\left (\theta_{\mathrm m} \right )} \right. \nonumber\\
&& + \left. \cos^{2}{\left (\frac{\theta}{2} \right )} \cos^{2}{\left (\theta_{\mathrm m} \right)}\right)
\end{eqnarray}
The critical points of (\ref{eq_av}) follow from $\nabla e = 0$. We are interested in the local minima and the saddle points.
Therefore we compute the discriminant
\begin{equation}
D = \frac{\partial^{2} e} {\partial \varphi_{\mathrm m}^{2}} \frac{\partial^{2}  e } { \partial \theta_{\mathrm m}^{2} } - 
\left( \frac{\partial^{2} e } { \partial \varphi_{\mathrm m} \partial \theta_{\mathrm m} }\right)^{2}
\end{equation}
and apply the conditions \cite{thomas1998} for a local minimum: $D > 0$ and $\frac{\partial^{2}  e} {\partial \varphi_{\mathrm m}^{2}} > 0$ and for a saddle point: $D < 0$.
Using the above conditions and $K<0$ in (\ref{eq_av}) we identify minima at $(\varphi_{\mathrm m}, \theta_{\mathrm m}) = (0,\pi/2),(\pi,\pi/2)$ and saddle points at $(\pi/2,\pi/2),(3\pi/2,\pi/2)$.{}
These critical points can also be found by visual inspection of Fig.~\ref{fig_ani}f. We see that the $x$-axis is an easy axis. To switch the magnetization from $+x$ to $-x$ we have to pass a saddle point. We define the height of the energy barrier as effective uniaxial anisotropy constant 
$K_{\mathrm{eff}} =  e(\pi/2,\pi/2) - e(0,\pi/2)$. The effective uniaxial anistropy constant as function of the angle between the easy planes is
\begin{equation}
\label{eq_keff}
K_{\mathrm{eff}} = -K \sin^{2}\frac{\theta}{2} = |K| \sin^{2}\frac{\theta}{2}
\end{equation}
This is equivalent to the effective anisotropy derived by Skomski \cite{skomski2016permanent} previously.
Fig.~\ref{fig_keff} gives the effective uniaxial anisotropy of two strongly exchange-coupled particles with easy-plane anisotropy.

Twinning by deformation \cite{partridge1967crystallography,britton2015mechanistic} might be one way to form a nanostructure with well defined angles between the crystallites. The angles used in Fig.~\ref{fig_ani}e to Fig.~\ref{fig_ani}h result from possible crystallite orientations between twins in Fe$_3$Sn. We used the software $Twiny$ \cite{nespolo2013} to compute relative orientations between the twins of hexagonal Fe$_3$Sn with lattice parameters $a=b=0.5456$~nm and $c=4.334$~nm \cite{osti_1196079}. Another way for forming structures with effective uniaxial anisotropy from easy-plane crystallites is field assisted cluster deposition.\cite{balasubramanian_high-coercivity_2016} Fig.~\ref{fig_ani}i shows a sketch of the particle cluster. Here the $c$ axes of the crystals are randomly oriented in the $yz$-plane. As a consequence intersection lines between the easy planes are all parallel to the $x$-axis which is the effective uniaxial anisotropy direction.   
In addition to the angle between the easy planes, the size of the exchange coupled crystallites is important. We used a finite element micromagnetic solver \cite{suess2002time} in order to compute the coercive field of the various structures as function of particle size. Since we are interested how the interplay between local easy-plane anisotropy and exchange interactions creates uniaxial behavior we switch off magnetostatic interactions in the simulations, in order to avoid shape effects. Fig.~\ref{fig_models} shows the structures used for the simulations: (a) two exchange-coupled particles, (b) a single grain with multiple twins, and (c) a cluster of 1000 particles generated by centroidal Voronoi tessellation \cite{quey2011large}.  The size of the features - particle size in (a) and (c) or layer thickness in (b) - was varied from 2~nm to 20~nm. For Fe$_{3}$Sn we use the anisotropy constant \cite{sales_ferromagnetism_2015}  $K = -1.8$~MJ/m$^3$, the magnetization \cite{sales_ferromagnetism_2015} $\mu_0 M_\mathrm{s}=1.48$~T and estimate the exchange constant $A = 10$~pJ/m. To compute the coercive field we solve the Landau-Lifshitz-Gilbert equation with a time varying external field. We change the field linearly from 0 to $-H_\mathrm{A} = - 2|K|/(\mu_0 M_\mathrm{s})$ in 3050~ns, which corresponds to a field rate of 1~mT/ns. The Gilbert damping constant was set to $\alpha = 1$. The initial state for each simulation was uniformly magnetized in the direction of the effective easy axis. To break the symmetry the external field is applied 0.1 degrees off the effective easy axis. The mesh size was $\delta_0/2 = 1.18$~nm. The Bloch parameter is defined as $\delta_0 = \sqrt{A/K}$. For structures (a) and (b) we used 90, 85, and 76 degrees as angles between the easy planes of neighboring crystals, which should give a reasonable effective uniaxial anisotropy according to Figures \ref{fig_ani}c, \ref{fig_ani}e, and \ref{fig_ani}f. 

\begin{table}
\caption{\label{tab_Keff} Effective uniaxial anisotropy for structures of Fe$_3$Sn and Sm$_{2}$Fe$_{14}$B. The line Fe$_3$Sn* contains the results computed with magnetostatic effects. The letters a, b, and c refer to the structures given in Fig.~\ref{fig_models}.}
\begin{tabular}{c c r r c}
\hline \hline
material & structure & angle($^o$)	& $s$(nm)	& $K_\mathrm{eff}$(MJ/m$^3$) \\ \hline
Fe$_3$Sn & a & 76 &  8 & 0.19 \\
Fe$_3$Sn & a & 85 &  8 & 0.22 \\
Fe$_3$Sn & a & 90 &  8 & 0.24 \\
Fe$_3$Sn & a & 76 & 12 & 0.10 \\
Fe$_3$Sn & a & 85 & 12 & 0.12 \\
Fe$_3$Sn & a & 90 & 12 & 0.15 \\
Fe$_3$Sn & b & 76 & 4 & 0.59 \\
Fe$_3$Sn* & b & 76 & 4 & 0.40 \\
Fe$_3$Sn & b & 76 & 10 & 0.31 \\
Fe$_3$Sn & b & 85 & 10 & 0.36 \\
Fe$_3$Sn & b & 90 & 10 & 0.39 \\
Fe$_3$Sn & c & random & 4 & 0.59 \\ 
Fe$_3$Sn & c & random & 10 & 0.23 \\ 
Sm$_{2}$Fe$_{14}$B & c & random & 4 & 1.24 \\ \hline  \hline
\end{tabular}
\end{table}

Fig.~\ref{fig_hc} shows the coercive field, $H_\mathrm{c}$, as function of the feature size, $s$, for the three different structures. The alternative axes give the coercive field and the feature size in multiples of the anisotropy field, $H_\mathrm{A}$, and the Bloch parameter, $\delta_0$, respectively. The coercive field decreases with increasing features size reaching $0.12 H_\mathrm{A}$ at $s = 3.4 \delta_0$ for two crystals at an angle of 85 degrees. The decrease of $H_\mathrm{c}$ with $s$ is less significant for the twinned grain. Since magnetostatic interactions were neglected, the Bloch parameter is the only characteristic length scale and the results in dimensionless units can be used to compute $H_\mathrm{c}(s)$ for materials other than Fe$_3$Sn. By equating the computed coercive field with the Stoner-Wohlfarth switching field \cite{stoner48}, we derived the effective uniaxial anisotropy for Fe$_3$Sn or Sm$_{2}$Fe$_{14}$B. The intrinsic properties of Sm$_{2}$Fe$_{14}$B used are: $K = 12$~MJ/m$^{3}$, $\mu_{0} M_{\mathrm s} = 1.49$~T, \cite{sagawa1985magnetic} and $A = 8.15$~pJ/m. The results are shown in Table~\ref{tab_Keff}. A local demagnetizing field may reduce the switching field of a structured particle. To investigate this effect, we computed an effective demagnetization factor as described by Fischbacher et al. \cite{fischbacher2017limits} for a grain with multiple twins (Fig.~\ref{fig_models}b) with $\theta = 76^{o}$. With an edge length of the cubic grain of 40 nm the effective demagnetization factor is $N_{\mathrm{eff}} = 0.24$. The computed coercive field is $\mu_{0}H_{\mathrm c}=0.67$~T. This gives an effective anisotropy of $K_\mathrm{eff} = 0.4$~MJ/m$^{3}$ (see line Fe$_{3}$Sn* in Table~\ref{tab_Keff}). Using these values, we estimate the energy density product of a magnet made of twinned Fe$_{3}$Sn grains to be 100~kJ/m$^{3}$ or higher.

In conclusion, our calculations explained how a moderate effective uniaxial anisotropy is induced in easy-plane materials by nanostructuring. This already has been achieved by magnetic-field-guided cluster deposition \cite{balasubramanian_high-coercivity_2016}. An additional method to create such structure might be deformation twinning.

Work supported by the NOVAMAG project, under Grant Agreement No. 686056, EU Horizon 2020 and the Austrian Science Fund (FWF): F4112 SFB ViCoM.

\bibliography{easy}

\providecommand{\noopsort}[1]{}\providecommand{\singleletter}[1]{#1}%
\begin{thebibliography}{23}%
\makeatletter
\providecommand \@ifxundefined [1]{%
 \@ifx{#1\undefined}
}%
\providecommand \@ifnum [1]{%
 \ifnum #1\expandafter \@firstoftwo
 \else \expandafter \@secondoftwo
 \fi
}%
\providecommand \@ifx [1]{%
 \ifx #1\expandafter \@firstoftwo
 \else \expandafter \@secondoftwo
 \fi
}%
\providecommand \natexlab [1]{#1}%
\providecommand \enquote  [1]{``#1''}%
\providecommand \bibnamefont  [1]{#1}%
\providecommand \bibfnamefont [1]{#1}%
\providecommand \citenamefont [1]{#1}%
\providecommand \href@noop [0]{\@secondoftwo}%
\providecommand \href [0]{\begingroup \@sanitize@url \@href}%
\providecommand \@href[1]{\@@startlink{#1}\@@href}%
\providecommand \@@href[1]{\endgroup#1\@@endlink}%
\providecommand \@sanitize@url [0]{\catcode `\\12\catcode `\$12\catcode
  `\&12\catcode `\#12\catcode `\^12\catcode `\_12\catcode `\%12\relax}%
\providecommand \@@startlink[1]{}%
\providecommand \@@endlink[0]{}%
\providecommand \url  [0]{\begingroup\@sanitize@url \@url }%
\providecommand \@url [1]{\endgroup\@href {#1}{\urlprefix }}%
\providecommand \urlprefix  [0]{URL }%
\providecommand \Eprint [0]{\href }%
\providecommand \doibase [0]{http://dx.doi.org/}%
\providecommand \selectlanguage [0]{\@gobble}%
\providecommand \bibinfo  [0]{\@secondoftwo}%
\providecommand \bibfield  [0]{\@secondoftwo}%
\providecommand \translation [1]{[#1]}%
\providecommand \BibitemOpen [0]{}%
\providecommand \bibitemStop [0]{}%
\providecommand \bibitemNoStop [0]{.\EOS\space}%
\providecommand \EOS [0]{\spacefactor3000\relax}%
\providecommand \BibitemShut  [1]{\csname bibitem#1\endcsname}%
\let\auto@bib@innerbib\@empty
\bibitem [{\citenamefont {Gutfleisch}\ \emph {et~al.}(2011)\citenamefont
  {Gutfleisch}, \citenamefont {Willard}, \citenamefont {Br{\"u}ck},
  \citenamefont {Chen}, \citenamefont {Sankar},\ and\ \citenamefont
  {Liu}}]{gutfleisch_magnetic_2011}%
  \BibitemOpen
  \bibfield  {author} {\bibinfo {author} {\bibfnamefont {O.}~\bibnamefont
  {Gutfleisch}}, \bibinfo {author} {\bibfnamefont {M.~A.}\ \bibnamefont
  {Willard}}, \bibinfo {author} {\bibfnamefont {E.}~\bibnamefont {Br{\"u}ck}},
  \bibinfo {author} {\bibfnamefont {C.~H.}\ \bibnamefont {Chen}}, \bibinfo
  {author} {\bibfnamefont {S.~G.}\ \bibnamefont {Sankar}}, \ and\ \bibinfo
  {author} {\bibfnamefont {J.~P.}\ \bibnamefont {Liu}},\ }\href@noop {}
  {\bibfield  {journal} {\bibinfo  {journal} {Adv. Mat.}\ }\textbf {\bibinfo
  {volume} {23}},\ \bibinfo {pages} {821} (\bibinfo {year} {2011})}\BibitemShut
  {NoStop}%
\bibitem [{\citenamefont {Drebov}\ \emph {et~al.}(2013)\citenamefont {Drebov},
  \citenamefont {Martinez-Limia}, \citenamefont {Kunz}, \citenamefont {Gola},
  \citenamefont {Shigematsu}, \citenamefont {Eckl}, \citenamefont {Gumbsch},\
  and\ \citenamefont {Els{\"a}sser}}]{drebov_ab_2013}%
  \BibitemOpen
  \bibfield  {author} {\bibinfo {author} {\bibfnamefont {N.}~\bibnamefont
  {Drebov}}, \bibinfo {author} {\bibfnamefont {A.}~\bibnamefont
  {Martinez-Limia}}, \bibinfo {author} {\bibfnamefont {L.}~\bibnamefont
  {Kunz}}, \bibinfo {author} {\bibfnamefont {A.}~\bibnamefont {Gola}}, \bibinfo
  {author} {\bibfnamefont {T.}~\bibnamefont {Shigematsu}}, \bibinfo {author}
  {\bibfnamefont {T.}~\bibnamefont {Eckl}}, \bibinfo {author} {\bibfnamefont
  {P.}~\bibnamefont {Gumbsch}}, \ and\ \bibinfo {author} {\bibfnamefont
  {C.}~\bibnamefont {Els{\"a}sser}},\ }\href@noop {} {\bibfield  {journal}
  {\bibinfo  {journal} {New J. Phys.}\ }\textbf {\bibinfo {volume} {15}},\
  \bibinfo {pages} {125023} (\bibinfo {year} {2013})}\BibitemShut {NoStop}%
\bibitem [{\citenamefont {Skomski}\ and\ \citenamefont
  {Coey}(2016)}]{skomski_magnetic_2016}%
  \BibitemOpen
  \bibfield  {author} {\bibinfo {author} {\bibfnamefont {R.}~\bibnamefont
  {Skomski}}\ and\ \bibinfo {author} {\bibfnamefont {J.}~\bibnamefont {Coey}},\
  }\href@noop {} {\bibfield  {journal} {\bibinfo  {journal} {Scr. Mater.}\
  }\textbf {\bibinfo {volume} {112}},\ \bibinfo {pages} {3} (\bibinfo {year}
  {2016})}\BibitemShut {NoStop}%
\bibitem [{\citenamefont {Brennan}\ \emph {et~al.}(1995)\citenamefont
  {Brennan}, \citenamefont {Skomski}, \citenamefont {Cugat},\ and\
  \citenamefont {Coey}}]{brennan_anisotropy_1995}%
  \BibitemOpen
  \bibfield  {author} {\bibinfo {author} {\bibfnamefont {S.}~\bibnamefont
  {Brennan}}, \bibinfo {author} {\bibfnamefont {R.}~\bibnamefont {Skomski}},
  \bibinfo {author} {\bibfnamefont {O.}~\bibnamefont {Cugat}}, \ and\ \bibinfo
  {author} {\bibfnamefont {J.}~\bibnamefont {Coey}},\ }\href@noop {} {\bibfield
   {journal} {\bibinfo  {journal} {J. Magn. Magn. Mater.}\ }\textbf {\bibinfo
  {volume} {140}},\ \bibinfo {pages} {971} (\bibinfo {year}
  {1995})}\BibitemShut {NoStop}%
\bibitem [{\citenamefont {Goll}\ \emph {et~al.}(2015)\citenamefont {Goll},
  \citenamefont {Loeffler}, \citenamefont {Herbst}, \citenamefont {Frey},
  \citenamefont {Goeb}, \citenamefont {Grubesa}, \citenamefont {Hohs},
  \citenamefont {Kopp}, \citenamefont {Pflanz}, \citenamefont {Stein},\ and\
  \citenamefont {Schneider}}]{goll_magnetic_2015}%
  \BibitemOpen
  \bibfield  {author} {\bibinfo {author} {\bibfnamefont {D.}~\bibnamefont
  {Goll}}, \bibinfo {author} {\bibfnamefont {R.}~\bibnamefont {Loeffler}},
  \bibinfo {author} {\bibfnamefont {J.}~\bibnamefont {Herbst}}, \bibinfo
  {author} {\bibfnamefont {C.}~\bibnamefont {Frey}}, \bibinfo {author}
  {\bibfnamefont {S.}~\bibnamefont {Goeb}}, \bibinfo {author} {\bibfnamefont
  {T.}~\bibnamefont {Grubesa}}, \bibinfo {author} {\bibfnamefont
  {D.}~\bibnamefont {Hohs}}, \bibinfo {author} {\bibfnamefont {A.}~\bibnamefont
  {Kopp}}, \bibinfo {author} {\bibfnamefont {U.}~\bibnamefont {Pflanz}},
  \bibinfo {author} {\bibfnamefont {R.}~\bibnamefont {Stein}}, \ and\ \bibinfo
  {author} {\bibfnamefont {G.}~\bibnamefont {Schneider}},\ }\href@noop {}
  {\bibfield  {journal} {\bibinfo  {journal} {Phys. Status Solidi RRL}\
  }\textbf {\bibinfo {volume} {9}},\ \bibinfo {pages} {603} (\bibinfo {year}
  {2015})}\BibitemShut {NoStop}%
\bibitem [{\citenamefont {Sales}\ \emph {et~al.}(2015)\citenamefont {Sales},
  \citenamefont {Saparov}, \citenamefont {McGuire}, \citenamefont {Singh},\
  and\ \citenamefont {Parker}}]{sales_ferromagnetism_2015}%
  \BibitemOpen
  \bibfield  {author} {\bibinfo {author} {\bibfnamefont {B.~C.}\ \bibnamefont
  {Sales}}, \bibinfo {author} {\bibfnamefont {B.}~\bibnamefont {Saparov}},
  \bibinfo {author} {\bibfnamefont {M.~A.}\ \bibnamefont {McGuire}}, \bibinfo
  {author} {\bibfnamefont {D.~J.}\ \bibnamefont {Singh}}, \ and\ \bibinfo
  {author} {\bibfnamefont {D.~S.}\ \bibnamefont {Parker}},\ }\href@noop {}
  {\bibfield  {journal} {\bibinfo  {journal} {Sci. Rep.}\ }\textbf {\bibinfo
  {volume} {4}} (\bibinfo {year} {2015})}\BibitemShut {NoStop}%
\bibitem [{\citenamefont {Skomski}, \citenamefont {Hadjipanayis},\ and\
  \citenamefont {Sellmyer}(2009)}]{skomski2009graded}%
  \BibitemOpen
  \bibfield  {author} {\bibinfo {author} {\bibfnamefont {R.}~\bibnamefont
  {Skomski}}, \bibinfo {author} {\bibfnamefont {G.}~\bibnamefont
  {Hadjipanayis}}, \ and\ \bibinfo {author} {\bibfnamefont {D.~J.}\
  \bibnamefont {Sellmyer}},\ }\href@noop {} {\bibfield  {journal} {\bibinfo
  {journal} {J. Appl. Phys.}\ }\textbf {\bibinfo {volume} {105}},\ \bibinfo
  {pages} {07A733} (\bibinfo {year} {2009})}\BibitemShut {NoStop}%
\bibitem [{\citenamefont {Balasubramanian}\ \emph {et~al.}(2016)\citenamefont
  {Balasubramanian}, \citenamefont {Manchanda}, \citenamefont {Skomski},
  \citenamefont {Mukherjee}, \citenamefont {Valloppilly}, \citenamefont {Das},
  \citenamefont {Hadjipanayis},\ and\ \citenamefont
  {Sellmyer}}]{balasubramanian_high-coercivity_2016}%
  \BibitemOpen
  \bibfield  {author} {\bibinfo {author} {\bibfnamefont {B.}~\bibnamefont
  {Balasubramanian}}, \bibinfo {author} {\bibfnamefont {P.}~\bibnamefont
  {Manchanda}}, \bibinfo {author} {\bibfnamefont {R.}~\bibnamefont {Skomski}},
  \bibinfo {author} {\bibfnamefont {P.}~\bibnamefont {Mukherjee}}, \bibinfo
  {author} {\bibfnamefont {S.~R.}\ \bibnamefont {Valloppilly}}, \bibinfo
  {author} {\bibfnamefont {B.}~\bibnamefont {Das}}, \bibinfo {author}
  {\bibfnamefont {G.~C.}\ \bibnamefont {Hadjipanayis}}, \ and\ \bibinfo
  {author} {\bibfnamefont {D.~J.}\ \bibnamefont {Sellmyer}},\ }\href@noop {}
  {\bibfield  {journal} {\bibinfo  {journal} {Appl. Phys. Lett.}\ }\textbf
  {\bibinfo {volume} {108}},\ \bibinfo {pages} {152406} (\bibinfo {year}
  {2016})}\BibitemShut {NoStop}%
\bibitem [{\citenamefont {Coey}(2012)}]{coey2012permanent}%
  \BibitemOpen
  \bibfield  {author} {\bibinfo {author} {\bibfnamefont {J.}~\bibnamefont
  {Coey}},\ }\href@noop {} {\bibfield  {journal} {\bibinfo  {journal} {Scr.
  Mater.}\ }\textbf {\bibinfo {volume} {67}},\ \bibinfo {pages} {524--529}
  (\bibinfo {year} {2012})}\BibitemShut {NoStop}%
\bibitem [{\citenamefont {Hirosawa}, \citenamefont {Nishino},\ and\
  \citenamefont {Miyashita}(2017)}]{hirosawa2017perspectives}%
  \BibitemOpen
  \bibfield  {author} {\bibinfo {author} {\bibfnamefont {S.}~\bibnamefont
  {Hirosawa}}, \bibinfo {author} {\bibfnamefont {M.}~\bibnamefont {Nishino}}, \
  and\ \bibinfo {author} {\bibfnamefont {S.}~\bibnamefont {Miyashita}},\
  }\href@noop {} {\bibfield  {journal} {\bibinfo  {journal} {Adv. Nat. Sci.:
  Nanosci. Nanotechnol.}\ }\textbf {\bibinfo {volume} {8}},\ \bibinfo {pages}
  {013002} (\bibinfo {year} {2017})}\BibitemShut {NoStop}%
\bibitem [{\citenamefont {Sagawa}\ \emph {et~al.}(1985)\citenamefont {Sagawa},
  \citenamefont {Fujimura}, \citenamefont {Yamamoto}, \citenamefont
  {Matsuura},\ and\ \citenamefont {Hirosawa}}]{sagawa1985magnetic}%
  \BibitemOpen
  \bibfield  {author} {\bibinfo {author} {\bibfnamefont {M.}~\bibnamefont
  {Sagawa}}, \bibinfo {author} {\bibfnamefont {S.}~\bibnamefont {Fujimura}},
  \bibinfo {author} {\bibfnamefont {H.}~\bibnamefont {Yamamoto}}, \bibinfo
  {author} {\bibfnamefont {Y.}~\bibnamefont {Matsuura}}, \ and\ \bibinfo
  {author} {\bibfnamefont {S.}~\bibnamefont {Hirosawa}},\ }\href@noop {}
  {\bibfield  {journal} {\bibinfo  {journal} {J. Appl. Phys.}\ }\textbf
  {\bibinfo {volume} {57}},\ \bibinfo {pages} {4094--4096} (\bibinfo {year}
  {1985})}\BibitemShut {NoStop}%
\bibitem [{\citenamefont {Herzer}(2005)}]{herzer2005}%
  \BibitemOpen
  \bibfield  {author} {\bibinfo {author} {\bibfnamefont {G.}~\bibnamefont
  {Herzer}},\ }in\ \href@noop {} {\emph {\bibinfo {booktitle} {Properties and
  {Applications} of {Nanocrystalline} {Alloys} from {Amorphous}
  {Precursors}}}},\ \bibinfo {editor} {edited by\ \bibinfo {editor}
  {\bibfnamefont {B.}~\bibnamefont {Idzikowski}}, \bibinfo {editor}
  {\bibfnamefont {P.}~\bibnamefont {Švec}}, \ and\ \bibinfo {editor}
  {\bibfnamefont {M.}~\bibnamefont {Miglierini}}}\ (\bibinfo  {publisher}
  {Springer},\ \bibinfo {year} {2005})\ p.~\bibinfo {pages} {15}\BibitemShut
  {NoStop}%
\bibitem [{\citenamefont {Skomski}\ and\ \citenamefont
  {Coey}(1993)}]{skomski1993giant}%
  \BibitemOpen
  \bibfield  {author} {\bibinfo {author} {\bibfnamefont {R.}~\bibnamefont
  {Skomski}}\ and\ \bibinfo {author} {\bibfnamefont {J.}~\bibnamefont {Coey}},\
  }\href@noop {} {\bibfield  {journal} {\bibinfo  {journal} {Phys. Rev. B}\
  }\textbf {\bibinfo {volume} {48}},\ \bibinfo {pages} {15812} (\bibinfo {year}
  {1993})}\BibitemShut {NoStop}%
\bibitem [{\citenamefont {Skomski}(2016)}]{skomski2016permanent}%
  \BibitemOpen
  \bibfield  {author} {\bibinfo {author} {\bibfnamefont {R.}~\bibnamefont
  {Skomski}},\ }in\ \href@noop {} {\emph {\bibinfo {booktitle} {Novel
  Functional Magnetic Materials}}}\ (\bibinfo  {publisher} {Springer},\
  \bibinfo {year} {2016})\ pp.\ \bibinfo {pages} {359--395}\BibitemShut
  {NoStop}%
\bibitem [{\citenamefont {Thomas}\ and\ \citenamefont
  {Finney}(1998)}]{thomas1998}%
  \BibitemOpen
  \bibfield  {author} {\bibinfo {author} {\bibfnamefont {G.~B.}\ \bibnamefont
  {Thomas}}\ and\ \bibinfo {author} {\bibfnamefont {R.~L.}\ \bibnamefont
  {Finney}},\ }\href@noop {} {\emph {\bibinfo {title} {Calculus and Analytic
  Geometry}}}\ (\bibinfo  {publisher} {Addison-Wesley Publishing Company,
  Reading, Massachusetts},\ \bibinfo {year} {1998})\BibitemShut {NoStop}%
\bibitem [{\citenamefont {Partridge}(1967)}]{partridge1967crystallography}%
  \BibitemOpen
  \bibfield  {author} {\bibinfo {author} {\bibfnamefont {P.}~\bibnamefont
  {Partridge}},\ }\href@noop {} {\bibfield  {journal} {\bibinfo  {journal}
  {Metall. Rev.}\ }\textbf {\bibinfo {volume} {12}},\ \bibinfo {pages} {169}
  (\bibinfo {year} {1967})}\BibitemShut {NoStop}%
\bibitem [{\citenamefont {Britton}, \citenamefont {Dunne},\ and\ \citenamefont
  {Wilkinson}(2015)}]{britton2015mechanistic}%
  \BibitemOpen
  \bibfield  {author} {\bibinfo {author} {\bibfnamefont {T.}~\bibnamefont
  {Britton}}, \bibinfo {author} {\bibfnamefont {F.}~\bibnamefont {Dunne}}, \
  and\ \bibinfo {author} {\bibfnamefont {A.}~\bibnamefont {Wilkinson}},\
  }\href@noop {} {\bibfield  {journal} {\bibinfo  {journal} {Proc. R. Soc. A}\
  }\textbf {\bibinfo {volume} {471}},\ \bibinfo {pages} {20140881} (\bibinfo
  {year} {2015})}\BibitemShut {NoStop}%
\bibitem [{\citenamefont {Nespolo}\ and\ \citenamefont
  {Iordache}(2013)}]{nespolo2013}%
  \BibitemOpen
  \bibfield  {author} {\bibinfo {author} {\bibfnamefont {M.}~\bibnamefont
  {Nespolo}}\ and\ \bibinfo {author} {\bibfnamefont {C.}~\bibnamefont
  {Iordache}},\ }\href@noop {} {\bibfield  {journal} {\bibinfo  {journal} {J.
  Appl. Cryst.}\ }\textbf {\bibinfo {volume} {46}},\ \bibinfo {pages} {801}
  (\bibinfo {year} {2013})}\BibitemShut {NoStop}%
\bibitem [{\citenamefont {Persson}(2015)}]{osti_1196079}%
  \BibitemOpen
  \bibfield  {author} {\bibinfo {author} {\bibfnamefont {K.}~\bibnamefont
  {Persson}},\ }\href@noop {} {\bibfield  {journal} {\bibinfo  {journal}
  {Materials Data on Fe3Sn (SG:194) by Materials Project}\ ,\ \bibinfo {pages}
  {doi:10.17188/1196079}} (\bibinfo {year} {2015})}\BibitemShut {NoStop}%
\bibitem [{\citenamefont {Suess}\ \emph {et~al.}(2002)\citenamefont {Suess},
  \citenamefont {Tsiantos}, \citenamefont {Schrefl}, \citenamefont {Fidler},
  \citenamefont {Scholz}, \citenamefont {Forster}, \citenamefont {Dittrich},\
  and\ \citenamefont {Miles}}]{suess2002time}%
  \BibitemOpen
  \bibfield  {author} {\bibinfo {author} {\bibfnamefont {D.}~\bibnamefont
  {Suess}}, \bibinfo {author} {\bibfnamefont {V.}~\bibnamefont {Tsiantos}},
  \bibinfo {author} {\bibfnamefont {T.}~\bibnamefont {Schrefl}}, \bibinfo
  {author} {\bibfnamefont {J.}~\bibnamefont {Fidler}}, \bibinfo {author}
  {\bibfnamefont {W.}~\bibnamefont {Scholz}}, \bibinfo {author} {\bibfnamefont
  {H.}~\bibnamefont {Forster}}, \bibinfo {author} {\bibfnamefont
  {R.}~\bibnamefont {Dittrich}}, \ and\ \bibinfo {author} {\bibfnamefont
  {J.}~\bibnamefont {Miles}},\ }\href@noop {} {\bibfield  {journal} {\bibinfo
  {journal} {J. Magn. Magn. Mater.}\ }\textbf {\bibinfo {volume} {248}},\
  \bibinfo {pages} {298} (\bibinfo {year} {2002})}\BibitemShut {NoStop}%
\bibitem [{\citenamefont {Quey}, \citenamefont {Dawson},\ and\ \citenamefont
  {Barbe}(2011)}]{quey2011large}%
  \BibitemOpen
  \bibfield  {author} {\bibinfo {author} {\bibfnamefont {R.}~\bibnamefont
  {Quey}}, \bibinfo {author} {\bibfnamefont {P.}~\bibnamefont {Dawson}}, \ and\
  \bibinfo {author} {\bibfnamefont {F.}~\bibnamefont {Barbe}},\ }\href@noop {}
  {\bibfield  {journal} {\bibinfo  {journal} {Comput. Methods Appl. Mech.
  Eng.}\ }\textbf {\bibinfo {volume} {200}},\ \bibinfo {pages} {1729} (\bibinfo
  {year} {2011})}\BibitemShut {NoStop}%
\bibitem [{\citenamefont {Stoner}\ and\ \citenamefont
  {Wohlfarth}(1948)}]{stoner48}%
  \BibitemOpen
  \bibfield  {author} {\bibinfo {author} {\bibfnamefont {E.~C.}\ \bibnamefont
  {Stoner}}\ and\ \bibinfo {author} {\bibfnamefont {E.}~\bibnamefont
  {Wohlfarth}},\ }\href@noop {} {\bibfield  {journal} {\bibinfo  {journal}
  {Philos. Trans. A Math. Phys. Eng. Sci.}\ }\textbf {\bibinfo {volume}
  {240}},\ \bibinfo {pages} {599} (\bibinfo {year} {1948})}\BibitemShut
  {NoStop}%
\bibitem [{\citenamefont {Fischbacher}\ \emph {et~al.}(2017)\citenamefont
  {Fischbacher}, \citenamefont {Kovacs}, \citenamefont {Oezelt}, \citenamefont
  {Gusenbauer}, \citenamefont {Schrefl}, \citenamefont {Exl}, \citenamefont
  {Givord}, \citenamefont {Dempsey}, \citenamefont {Zimanyi}, \citenamefont
  {Winklhofer}, \citenamefont {Hrkac}, \citenamefont {Chantrell}, \citenamefont
  {Sakuma}, \citenamefont {Yano}, \citenamefont {Kato}, \citenamefont {Shoji},\
  and\ \citenamefont {Manabe}}]{fischbacher2017limits}%
  \BibitemOpen
  \bibfield  {author} {\bibinfo {author} {\bibfnamefont {J.}~\bibnamefont
  {Fischbacher}}, \bibinfo {author} {\bibfnamefont {A.}~\bibnamefont {Kovacs}},
  \bibinfo {author} {\bibfnamefont {H.}~\bibnamefont {Oezelt}}, \bibinfo
  {author} {\bibfnamefont {M.}~\bibnamefont {Gusenbauer}}, \bibinfo {author}
  {\bibfnamefont {T.}~\bibnamefont {Schrefl}}, \bibinfo {author} {\bibfnamefont
  {L.}~\bibnamefont {Exl}}, \bibinfo {author} {\bibfnamefont {D.}~\bibnamefont
  {Givord}}, \bibinfo {author} {\bibfnamefont {N.}~\bibnamefont {Dempsey}},
  \bibinfo {author} {\bibfnamefont {G.}~\bibnamefont {Zimanyi}}, \bibinfo
  {author} {\bibfnamefont {M.}~\bibnamefont {Winklhofer}}, \bibinfo {author}
  {\bibfnamefont {G.}~\bibnamefont {Hrkac}}, \bibinfo {author} {\bibfnamefont
  {R.}~\bibnamefont {Chantrell}}, \bibinfo {author} {\bibfnamefont
  {N.}~\bibnamefont {Sakuma}}, \bibinfo {author} {\bibfnamefont
  {M.}~\bibnamefont {Yano}}, \bibinfo {author} {\bibfnamefont {A.}~\bibnamefont
  {Kato}}, \bibinfo {author} {\bibfnamefont {T.}~\bibnamefont {Shoji}}, \ and\
  \bibinfo {author} {\bibfnamefont {A.}~\bibnamefont {Manabe}},\ }\href@noop {}
  {\bibfield  {journal} {\bibinfo  {journal} {Appl. Phys. Lett.}\ }\textbf
  {\bibinfo {volume} {111}},\ \bibinfo {pages} {072404} (\bibinfo {year}
  {2017})}\BibitemShut {NoStop}%
\end{thebibliography}%

\end{document}